\documentclass{aastex}
\usepackage{emulateapj5,epsfig,onecolfloat,apjfonts}   

\shorttitle{Power Spectrum Normalization from SDSS/RASS and REFLEX}
\shortauthors{P.~T.~P.~Viana, R.~C.~Nichol and
A.~R.~Liddle}

\begin{document}

\twocolumn
[
\title{Constraining the Matter Power Spectrum Normalization using the
SDSS/RASS and REFLEX Cluster surveys}

\author{Pedro T. P.  Viana} 
\affil{Centro de Astrof\'{\i}sica, 
Universidade do Porto, Rua das Estrelas, 4150-762 Porto, Portugal\\
Departamento de Matem\'{a}tica Aplicada, Faculdade de Ci\^{e}ncias, 
Universidade do Porto, Rua do Campo Alegre, 687, 4169-007 Porto, Portugal\\~}

\author{Robert C.~Nichol}
\affil{Physics Department, Carnegie Mellon
University, 5000 Forbes Avenue, Pittsburgh PA15213, U.S.A.\\~}

\author{Andrew R.~Liddle}
\affil{Astronomy Centre, University of Sussex, Falmer, Brighton BN1 9QJ, UK}

\begin{abstract}
We describe a new approach to constrain the amplitude of the power spectrum of 
matter perturbations in the Universe, parametrized by $\sigma_8$ as a function 
of the matter density $\Omega_0$. We compare the galaxy 
cluster X-ray luminosity function of the REFLEX survey with the theoretical 
mass function of Jenkins et al.~(2001), using the mass--luminosity relationship 
obtained from weak lensing data for a sample of galaxy clusters 
identified in Sloan Digital Sky Survey commissioning data and confirmed 
through cross-correlation with the {\it ROSAT} all-sky survey. We find  
$\sigma_{8} = 0.38 \, \Omega_{0}^{-0.48+0.27\Omega_{0}}$, which is significantly
different from most previous results derived from comparable calculations that 
used the X-ray temperature function. We discuss possible sources of 
systematic error that may cause such a discrepancy, and in the process 
uncover a possible inconsistency between the REFLEX luminosity 
function and the relation between cluster X-ray luminosity and mass 
obtained by Reiprich \& B\"{o}hringer (2001). 
\end{abstract}

]

\section{Introduction}

The present-day number density of galaxy clusters remains one of the most
powerful constraints on the amplitude of matter perturbations in
the Universe.  This is usually defined in terms of $\sigma_{8}$, which is the
dispersion of the mass field smoothed on a scale of $8\,h^{-1}\;{\rm Mpc}^{-1}$,
where $h$ is the present value of the Hubble parameter in units of $100\;{\rm
km}\,{\rm s}^{-1}\,{\rm Mpc}^{-1}$.  Most often the local X-ray cluster
temperature function is used for this purpose, given that the X-ray temperature
has been the best observable from which to estimate cluster mass.  While
observation can readily give the X-ray temperature of clusters, theory
can only easily predict the cluster mass function.  To bridge the gap,
theoretical modelling of clusters is used to provide a relation between mass and
temperature, which in the most sophisticated treatments is taken to depend on
both redshift and the underlying cosmology.

However, a drawback in the use of the local X-ray cluster temperature function
is that only a few tens of clusters have had their X-ray temperature estimated.  
Consequently, authors using samples that only partially
overlap have obtained significantly different cluster temperature functions, 
and thus estimates for $\sigma_{8}$ (e.g.~compare Eke, Cole \& Frenk 1996; 
Henry 1997, 2000; Markevitch 1998; Blanchard et al. 2000; Pierpaoli et 
al.~2001).  
Statistically it would be more robust if the number density of local clusters 
could be estimated from much larger samples.

We describe here a new approach which avoids working with the X-ray
temperature function. We will instead use the luminosity function from 
the {\it ROSAT}-ESO Flux Limited X-Ray (REFLEX) galaxy cluster survey 
(B\"{o}hringer et al.~2001a,b), which contains 452 clusters, 
to estimate the local cluster number
density. We relate X-ray luminosity to cluster mass by taking advantage
of a weak shear lensing analysis (Sheldon et al.~2001) of a sample of 42 galaxy
clusters identified in data from the commissioning phase of the Sloan Digital
Sky Survey (SDSS), and cross-checked via correlation with the {\it ROSAT}
All-Sky Survey (RASS) (Nichol et al.~2001). This relation is then used to 
compare the REFLEX luminosity function with the cluster mass function, 
leading to the estimation of $\sigma_{8}$, without any need to model the 
cluster mass--temperature relation (as in e.g.~Viana \& Liddle 1996, 1999).

\section{Methodology}

We consider spatially-flat cosmological models, where part of the energy
density may be due to a cosmological constant, containing a spectrum of
primordial adiabatic density perturbations.  This family includes the current
standard cosmological model, with a present-day matter density $\Omega_{0}\simeq
0.3$, which provides the best fit to the full compilation of structure formation
data (e.g.~Durrer \& Novosyadlyj 2001; Wang, Tegmark \& Zaldarriaga 2001).  We
take the present-day shape of the matter power spectrum to be
well approximated by that of a cold dark matter model with scale-invariant
primordial density perturbations and shape parameter $\Gamma$ in the range 
[0.08,0.28]; this interval is the average of the best-fit values coming from 
the preliminary analysis of the 2dF (Percival et al.~2001) and SDSS 
(Dodelson et al.~2001; Szalay et al.~2001) data, taking into account both 
statistical and systematic uncertainties. We will assume that $\Gamma$ has 
an equal probability of taking any value within the interval given. Note, 
however, that ultimately the local cluster number density depends 
significantly only on $\sigma_8$ and $\Omega_{0}$.

The expected halo mass function for each set of cosmological parameters is 
estimated via the fitting function of Jenkins et al.~(2001), obtained 
by analysing data assembled from various large $N$-body simulations. Following 
White~(2001) we consider the halo mass to be that given by the virial relation. 
The high-mass end of the halo mass function is better estimated using the 
Jenkins et al.~(2001) result rather than the usual
Press--Schechter ansatz (Press \& Schechter 1974). The latter leads to a smaller
number of high-mass halos at fixed $\sigma_8$, thus requiring
systematically higher values of $\sigma_8$, by about eight percent, in order for 
the local cluster abundance to be reproduced. We cautiously model the uncertainty 
in the normalization of the mass function by means of a Gaussian distribution 
with a 10 percent dispersion around the mean 
(see Fig. 8 of Jenkins et al.~2001).

The local cluster number density was obtained by integrating the REFLEX X-ray
luminosity function in the [0.1,2.4] keV band upwards from the mean luminosity
of the 42 clusters in the SDSS/RASS sample, which is 
$(0.17\pm0.03)\times10^{44}\;h^{-2}\,{\rm erg/s}$.  Their average redshift 
is 0.1, which roughly coincides with the mean redshift of the REFLEX clusters 
with similar or higher luminosity.  For the 42 SDSS/RASS clusters, Sheldon et
al.~(2001) obtained, via statistical weak lensing analysis, a mean projected
mass within $r_{500}$ of
$(0.9\pm0.2)\times10^{14}\;h^{-1}\,M_{\odot}$, where $r_{500}$ is the radius 
within 
which
the cluster mean density falls to 500 times the critical density at the redshift
of observation.  Following Sheldon et al.~(2001) we assume
the cluster density profile to behave like that of a singular isothermal 
sphere, i.e.~$\rho(r)\propto r^{-2}$.  We can now calculate the mean cluster
mass within the three-dimensional radius $r_{500}$, and then convert this to a
mean virial mass.  Taking this mass as the lower limit in the integration of the
Jenkins et al.~(2001) mass function will then yield the expected local cluster
number density for clusters more luminous than the mean of the SDSS/RASS sample,
as a function of $\sigma_{8}$ and $\Omega_{0}$.  By comparison with the REFLEX
estimate, best-fit values for $\sigma_{8}$ as a function of $\Omega_{0}$ can
then be obtained.

\section{Results}

Table 1 shows the 95 per cent confidence interval on $\sigma_8$ 
obtained using the REFLEX X-ray luminosity function and the relation between 
cluster mass and X-ray luminosity for the full SDSS/RASS sample. This interval 
was determined via Monte Carlo simulations, which 
incorporated all the uncertainties previously mentioned that affect the present 
estimation of $\sigma_8$. We find that the most probable value for $\sigma_8$ 
can be accurately represented by the fitting function
\begin{equation}
\sigma_{8} = 0.38 \; \Omega_{0}^{-0.48+0.27\Omega_{0}} \,,
\end{equation}
with a 95 per cent uncertainty around 15\%.

This result is significantly lower than, and barely compatible with, most 
determinations of $\sigma_8$ based on the local cluster X-ray temperature
function (e.g.~compare with Eke, Cole \& Frenk 1996; Henry 1997, 2000; 
Viana \& Liddle 1999; Blanchard et al.~2000; Pierpaoli et al.~2001), with
the exception of Seljak (2001). His analysis differs from the others in 
that he used the relation between cluster temperature and mass derived by 
Finoguenov, Reiprich \& B\"{o}hringer (2001) from X-ray data, rather 
than one obtained from hydrodynamical $N$-body simulations. The earlier 
work of Markevitch (1998) had already hinted at lower values for 
$\sigma_8$, at least in the case of $\Omega_{0}\simeq0.3$, if actual 
X-ray data was used to relate cluster temperature to mass. Similar 
results have been reached by Borgani et al. (2001), based on 
{\it ROSAT} Deep Cluster Survey (RDCS) data and 
the observed cluster X-ray temperature to luminosity relation, and 
Reiprich \& B\"{o}hringer (2001), by means of an empirical cluster 
mass function derived using X-ray data from a large cluster sample.

Other methods of measuring $\sigma_8$ lead to conflicting results. While 
high-redshift Lyman-$\alpha$ forest analyses (Croft et al.~1999, 2000; 
McDonald et al.~2000), support our findings, estimates based on cosmic shear 
data (H\"{o}kstra, Yee \& Gladders 2001; Maoli et al.~2001; 
Van Waerbeke et al.~2001) tend to favour higher values for $\sigma_8$ than 
those obtained here. Also, the 2dF galaxy survey, when combined with 
measurements 
of the amplitude of temperature anisotropies in the cosmic microwave background 
radiation (Efstathiou et al.~2001; Lahav et al.~2001), seems to require a value 
for $\sigma_8$ close to what we have found. 

\begin{table}[t]
\caption{Matter power spectrum normalization using the REFLEX 
luminosity function and the full SDSS/RASS sample, shown with 95\% confidence 
error bars.
\label{table1} 
}
\centering
\begin{tabular}{|c|c|c|}
\hline
$\Omega_0$ & $\sigma_8$\\
\hline
 1.0  &     $0.38\pm0.05$\\
 0.9  &     $0.39\pm0.06$\\
 0.8  &     $0.41\pm0.06$\\
 0.7  &     $0.43\pm0.06$\\
 0.6  &     $0.46\pm0.07$\\
 0.5  &     $0.49\pm0.07$\\
 0.4  &     $0.54\pm0.08$\\
 0.3  &     $0.61\pm0.10$\\
 0.2  &     $0.74\pm0.14$\\
 0.1  &     $1.09^{+0.33}_{-0.23}$\\
\hline
\end{tabular}
\end{table}

In order to test whether the low values obtained for $\sigma_8$ could be due to
hidden systematic errors in the weak lensing method used for cluster mass
estimation, we repeated the calculation of $\sigma_{8}$ for the two SDSS/RASS 
sub-samples discussed within Sheldon et al.~(2001), namely 
\begin{enumerate}
\item The 27 clusters with the lowest X-ray luminosities, on average 
$(0.09\pm0.02)\times10^{44}\;h^{-2}\,{\rm erg/s}$. They have a mean
redshift of 0.09 and a mean projected mass within $r_{500}$ of $M_{500}(1) =
(0.7 \pm 0.2) \times 10^{14} \; h^{-1} \, M_{\odot}$.
\item The 15 clusters with the highest X-ray luminosities, on average 
$(0.51\pm0.04)\times10^{44}\;h^{-2}\,{\rm erg/s}$. They have a mean 
redshift of 0.17 and a mean projected mass within $r_{500}$ of $M_{500}(2) = 
2.7^{+0.9}_{-1.1} \times 10^{14} \; h^{-1} \, M_{\odot}$.
\end{enumerate}  
Given that most of the REFLEX clusters with similar luminosities to those in
the SDSS/RASS sample have a redshift between 0.05 and 0.2 (B\"{o}hringer et
al.~2001a), with the higher luminosity clusters typically being at higher
redshifts, we will assume that the REFLEX luminosity function provides a good
representation of the underlying cluster luminosity function over this
redshift interval.  This is supported by an analysis of the Brightest Cluster
Survey (BCS), which showed that there is no strong evidence for evolution in
the cluster luminosity function out to at least $z=0.2$ (Ebeling et al.~1997).

Surprisingly, the most probable values for $\sigma_{8}$ according to each of
these two sub-samples are rather different. Sub-sample~1 yields
$\sigma_{8}=0.37\,\Omega_{0}^{-0.52+0.32\Omega_{0}}$, almost indistinguishable
from the result obtained from the full SDSS/RASS sample, while sub-sample~2 
gives $\sigma_{8} = 0.50\,\Omega_{0}^{-0.47+0.24\Omega_{0}}$ 
which is substantially higher.  This latter result is much more in line
with standard $\sigma_{8}$ estimates from the local cluster X-ray 
temperature function.

\section{Discussion}

In this section, we discuss our results and present several tests 
to assess their robustness. First, to determine whether our
results indicate an internal inconsistency in the weak lensing analysis of
Sheldon et al.~(2001), we performed Monte Carlo simulations where we used the
data from sub-sample 1 to compute the inferred mean projected masses within
$r_{500}$ for the clusters in sub-sample 2 and vice-versa, for various values
of $\Omega_0$.  We included all the uncertainties previously mentioned in the
simulations.  We found that, provided $\Omega_{0}>0.3$, sub-sample 1 implies a
value for $M_{500}$(2) at least as large as that deduced by Sheldon et
al.~(2001) less than 5 per cent of the time, while sub-sample 2 implies a value 
for $M_{500}$(1) as small as that
deduced by Sheldon et al.~(2001) less than 10 per cent
of the time.  For values of $\Omega_{0}$ between 0.1 and
0.3, the discrepancy decreases with $\Omega_{0}$, but not significantly.  We
therefore conclude that, within the context of the cosmological models we
discuss, the mean projected masses for the two SDSS/RASS cluster sub-samples
presented in Sheldon et al.~(2001) are only barely compatible within the
uncertainties associated with the estimation of $\sigma_{8}$.

This inconsistency could result from the assumption 
that the cluster mass profile is that of a singular isothermal 
sphere (SIS), which was motivated by the fact that 
within the radius for which there was weak
lensing data, the shear profile was close to that of a projected SIS.
Although observationally the issue is still unresolved (e.g.~Irwin, Bregman \&
Evrard 1999; Irwin \& Bregman 2000; White 2000; De Grandi \& Molendi 2001;
Komatsu \& Seljak 2001), it has become clear that $N$-body simulations produce
clusters which on average have outer mass profiles, up to the cluster virial
radius, that behave more like $\rho(r)\propto r^{-2.5}$ (Navarro, Frenk \&
White 1995, 1996, 1997; Thomas et al.~1998, 2001; Tittley \& Couchman 2001).
Consequently, we investigated the effect on the virial mass to
X-ray luminosity relation of assuming an outer cluster mass profile
parametrized by $\rho(r)\propto r^{-\beta}$.  We varied $\beta$ between 2 
(SIS case) and 3, and, as before, we took the mean projected masses
within $r_{500}$ to be those given by Sheldon et al.~(2001).  This translates 
to fixing the inner projected cluster masses to
be $M_{500}$(1) within $0.39 \, h^{-1} \; {\rm Mpc}$ for sub-sample 1,
$M_{500}$(2) within $0.57 \, h^{-1} \; {\rm Mpc}$ for sub-sample 2, and $0.9
\pm 0.2 \times 10^{14} \, M_{\odot}$ within $0.41 \, h^{-1} \; {\rm Mpc}$ for
the full SDSS/RASS sample (Sheldon et al.~2001).  As expected, we find that as
$\beta$ increases, the cluster virial mass decreases, hence lowering
$\sigma_8$.  This change is greater for smaller $\Omega_{0}$.  Consequently,
we find that the value for $\sigma_{8}$ derived using the Sheldon et
al.~(2001) and REFLEX data is barely affected by assuming different outer mass
profiles if $\Omega_{0}$ is close to one, whereas for low values of
$\Omega_{0}$ assuming a outer mass profile different from the true one may
introduce a significant systematic error in the calculation of $\sigma_{8}$.
For example, for $\beta=3$ we obtain
$\sigma_{8}=0.37\,\Omega_{0}^{-0.41+0.24\Omega_{0}}$ as the most probable
value that results from the weak lensing analysis for the full SDSS/RASS
sample.  However, it turns out that changing the outer cluster mass density
profile does not significantly mitigate the discrepancy between the values
obtained for $\sigma_{8}$ using the two SDSS/RASS sub-samples, even for low
values of $\Omega_{0}$.

The discrepancy between the $\sigma_{8}$ results obtained for the two
Sheldon et al.~(2001) sub-samples could be alleviated if the mean projected 
mass of the high-luminosity sample is overestimated, or that of the 
low-luminosity sample underestimated. The first hypothesis is much more 
probable, and could be due to a contribution to the mean projected mass 
by filamentary material infalling into the clusters along the line of 
sight (Cen 1997; Metzler et al. 1999; Reblinsky \& Bartelmann 1999; 
Metzler, White \& Loken 2001). Alternatively, the mean X-ray luminosity 
of the high-luminosity sample may have been underestimated, or that of the 
low-luminosity sample overestimated. The latter could be caused by AGN 
contamination, but the initial study of Miller et al. (2002, in preparation) 
indicates that this is not a serious problem in the case of the SDSS/RASS 
cluster survey. Furthermore, the observed relationship 
between X-ray cluster luminosity and measured velocity dispersion 
(from the SDSS spectroscopic sample) for both the Sheldon et al. (2001) 
and the Miller et al. (2002, in prep.) samples are in excellent agreement with 
the $L_{\rm X}$--$\sigma_{\rm v}$ relation of Mahdavi \& Geller (2001). 

Reiprich \& B\"{o}hringer (2001) used {\it ROSAT} and {\it ASCA} X-ray data on
106 clusters to obtain the relation between X-ray luminosity in the
[0.1,2.4] keV band and cluster mass in the form of $M_{500}$, with the cluster
masses estimated assuming hydrostatic equilibrium (Finoguenov et al.~2001; 
Reiprich \& B\"{o}h\-ringer 2001). They found
$L_{X}=10^{-18.59\pm1.23}\times M_{500}^{1.575\pm0.084}$ where $L_{X}$ is in
units of $10^{40}\,{\rm erg/s}$ and $M_{500}$ in units of $M_{\odot}$ (assuming
$h=0.5$). Substituting the mean luminosities for the SDSS/RASS full sample and
two sub-samples into this relation, one finds that the Sheldon et al.~(2001)
estimates for $M_{500}$ are well within the (extremely wide) range of possible
values. Conversely, the relation between $L_{X}$ and $M_{500}$ can be estimated 
by combining the Sheldon et al.~(2001) results with the shape of the cluster 
mass function, assumed well described by that of Jenkins et al.~(2001), 
and the data on the cluster luminosity function from the REFLEX
survey (B\"{o}hringer et al.~2001a,b).  Assuming the $L_{X}-M_{500}$ relation 
to be a power-law, we found that the normalization is essentially defined by the
Sheldon et al.~(2001) data, as expected, while the exponent is mainly governed
by the relative shape of the mass and luminosity (cumulative) functions.  Taking
into account all the uncertainties involved in the normalization of the Jenkins
et al.~(2001) mass function by means of the Sheldon et al.~(2001) data for the
full SDSS/RASS sample, and those associated with the REFLEX luminosity function,
we obtained through Monte Carlo simulations a 95 per cent confidence interval of
[1.6,3.7] for the exponent of the $L_{X}$-$M_{500}$ relation.  The allowed
interval does not change significantly if the data for either of the two SDSS/RASS
sub-samples is used instead to normalize the mass function. The 
low-luminosity sample yields [1.8,3.2], while from the high-luminosity 
sample we
get [1.5,3.1].  In any case, surprisingly, the preferred value is close to 2.2,
substantially higher than the $1.58\pm0.08$ at $1\sigma$ found by Reiprich \&
B\"{o}hringer (2001).  This analysis was performed for a flat universe with
$\Omega_{0}=0.3$, though very similar results were found for $\Omega_{0}=1$.  
In order to confirm that the exponent of the $L_{X}-M_{500}$ relation
is only weakly determined by the normalization of the mass function, and thus by
the Sheldon et al.~(2001) data, we varied $\sigma_{8}$ between 0.5 and 1.2 
(for $\Omega_{0}=0.3$) and found that the preferred value for the
exponent changes from 2.4 to 1.7.  Therefore, assuming the Jenkins et al.~(2001)
mass function provides an accurate description of the cluster mass function, the
discrepancy just found on the exponent of the $L_{X}-M_{500}$ relation means
that such a relation as obtained in Reiprich \& B\"{o}hringer (2001) is at best
only marginally consistent with the Sheldon et al.~(2001) data taken together
with the REFLEX luminosity function. And only if $\sigma_{8}$ is at the higher
end of recent estimates (see e.g.~Viana \& Liddle 1999) can the Reiprich \&
B\"{o}hringer (2001) $L_{X}-M_{500}$ relation be made consistent
with the REFLEX luminosity function, within the context of the cosmological
models discussed in this paper.

In calculating $\sigma_{8}$ we assumed that, on average, a galaxy cluster with
the mean X-ray luminosity of the SDSS/RASS full sample has $M_{500}$ equal to
the value estimated in Sheldon et al.~(2001) for such a sample.  This assumption
is prone to several biases, one being that, assuming that there is some
dispersion in the associated luminosity, the clusters of a given mass that
preferentially end up in a sample selected in the manner of the SDSS/RASS are the
most luminous ones.  This bias leads to an underestimation of the correct mass
corresponding to a given luminosity, and consequently to an underestimation of
$\sigma_{8}$, though it is difficult to say by how much.  The same effect takes
place due to another, more subtle, type of bias, arising from the fact that the
$L_{X}$--$M_{500}$ relation is not linear.  As a result, for a given sample the
mean $M_{500}$ is not proportional to the mean $L_{X}$.  Given that
$M_{500}\propto L_{X}^{n}$ with $n<1$, assuming that a cluster with the mean 
luminosity has a mass $M_{500}$ equal to the sample mean underestimates 
$M_{500}$ for such a cluster (because actually the mean
$M_{500}$ is proportional to the mean of $L_{X}^{n}$).  We have found that, for 
the luminosity dispersion of the
clusters in the SDSS/RASS full sample, the most probable value of $M_{500}$ for
a cluster with the mean $L_{X}$ may be underestimated by around
20 per cent.  This percentage is robust to changes in the assumed value for the
exponent $n$ (between 0.3 and 0.7) and to the possibility of
dispersion in the $L_{X}$--$M_{500}$ relation.  This bias leads to a possible
underestimation of $\sigma_{8}$ close to 10 per cent, almost independent of
$\Omega_{0}$.  However, given that the mean $M_{500}$ for the SDSS/RASS sample
is in fact not a mean of several independently calculated $M_{500}$, but the
result of a mean shear profile, and that the mean $L_{X}$ is weighted by each
cluster contribution to that profile, the above considerations may not be
directly applicable to the case in hand.  Note that because
the dispersion in luminosities is smaller for either of the two SDSS/RASS
sub-samples as compared to the full sample, the underestimation in 
$\sigma_{8}$ is smaller when it is estimated from the sub-samples, being 
closer to 5 per cent.  Unfortunately, none of these biases seem to be able to 
significantly narrow the discrepancy between the $\sigma_{8}$ values derived 
using each SDSS/RASS sub-sample.

\section{Conclusions}

We have applied a new approach for constraining the
normalization of the matter power spectrum, using 
the REFLEX X-ray luminosity function and the 
relation between cluster X-ray luminosity and mass 
obtained through weak lensing data from a preliminary small 
sample of SDSS/RASS clusters. We obtained $\sigma_{8}$ values significantly 
lower than other estimates based on cluster abundance data, with the 
exception of the recent results by 
Borgani et al.~(2001), Reiprich \& B\"{o}hringer (2001) and Seljak (2001).
However, systematic biases may affect our analysis, given that
barely consistent results are obtained when using subsets of the weak lensing
data.  This may be due to the small sample of 
clusters used, or an artifact of the techniques used in Sheldon et al.~(2001) 
for co-adding clusters to produce an ensemble averaged weak lensing 
signal. In the process, we found that comparing the REFLEX luminosity function 
and the Jenkins et al.~(2001) mass function implies that the relation between 
X-ray luminosity and cluster mass may be significantly steeper than previously 
thought. 

The SDSS/RASS data set we have used will be dwarfed by the final SDSS/RASS
catalogue (see Nichol et al.~2001), and surveys using the
{\it XMM-Newton} satellite should supply much greater information on cluster
luminosities (e.g.~Romer et al. 2001). The prospect of considerably improving
the constraint on $\sigma_{8}$ using this approach in the future is therefore
great. 
 
\acknowledgments

A.R.L.~was supported in part by the Leverhulme Trust. We thank Hans 
B\"{o}hringer, Bob Mann and Kathy Romer for useful discussions and comments, 
Erin Sheldon for providing the mean X-ray luminosities 
of the SDSS/RASS samples in the {\it ROSAT} [0.1,2.4] keV band, and Vincent 
Eke for pointing out the possible bias in the $\sigma_{8}$ estimation 
due to the non-linear nature of the $L_{X}-M_{500}$ relation. 



\end{document}